\documentclass[11pt]{article}
\usepackage{cospar}
\usepackage{url}

\usepackage{graphicx}
\usepackage[figuresright]{rotating}

\title{XMM-NEWTON OBSERVATIONS OF THE FIELD OF GAMMA-RAY BURST 980425}

\author{E. Pian\address{INAF, Osservatorio Astronomico di Trieste, Via G.B. Tiepolo 11, I-34131
Trieste, Italy},
P. Giommi\address{ASI Science Data Center, c/o ESRIN, Via G. Galilei, I-00044 Frascati, Italy},
L. Amati\address{IASF-CNR, Sezione di Bologna, Via P. Gobetti 101, I-40129 Bologna, Italy}, 
E. Costa\address{IASF-CNR, Sezione di Roma, via Fosso Del Cavaliere 100, I-00133 Roma, Italy}, 
J. Danziger$^1$,
M. Feroci$^4$,
M.T. Fiocchi$^2$,
F. Frontera\address{Physics Department, University of Ferrara, Via Paradiso 11, I-44100 Ferrara,
Italy}, 
C.~Kouveliotou\address{NASA MSFC, SD-50, Huntsville, AL 35812, USA}, 
N. Masetti$^3$,
L. Nicastro\address{IASF-CNR, Sezione di Palermo, via U. La Malfa 153, I-90146 Palermo, Italy},
E. Palazzi$^3$,
L. Piro$^4$, 
M. Tavani$^4$, 
and J.J.M.~in~'t~Zand\address{Space Research Organization Netherlands, Sorbonnelaan 2, 3584 CA
Utrecht, The Netherlands}}

\begin{document}

\maketitle

\begin{abstract}

The error box of GRB980425 has been observed by XMM-{\it Newton} in March 2002, with the aim
of measuring the late epoch X-ray emission of the supernova 1998bw and of clarifying its
supposed association with the GRB itself.  We present here the preliminary results obtained
with the EPIC PN camera. Our observations confirm the association between SN~1998bw and
GRB980425.  The EPIC PN measurement of the SN~1998bw flux is significantly below the
extrapolation of the power-law temporal trend fitted to the BeppoSAX points and implies a
faster temporal decay.  We propose different physical interpretations of the SN X-ray light
curve, according to whether it is produced by one or more radiation components.

\end{abstract}

\section*{INTRODUCTION}

The hypothesis that supernovae are the progenitors of Gamma-Ray Bursts (GRB) dates back to the
epoch of first GRB discovery (Colgate 1974) and has received support in recent years from the
detection of supernova features in the optical afterglows of GRBs. These are re-brightenings
at rest-frame intervals of 10-15 days after the GRB (e.g., Galama et al. 2000; Bloom et al.
2002; Price et al. 2003; Masetti et al. 2003);  circumburst media with wind characteristics
(Berger et al. 2001; Jaunsen et al. 2001; Price et al. 2002);  iron emission lines (Piro et
al. 1999; Reeves et al. 2002; Butler et al. 2003);  association of GRBs with star-forming
regions (Fruchter et al. 1999; Frail et al. 2002).  The most tempting hint of association
between GRBs and SNe is obviously the similarity of the intrinsic energy of these phenomena,
when collimation and beaming are taken into account in GRBs (e.g., Frail et al. 2001).

While these circumstances represent only possible evidences of a GRB-SN connection, on two
occasions a clear association has been found. GRB980425 and SN~1998bw have been detected
within very tight temporal and angular limits: they exploded simultaneously (with an
uncertainty of $\pm 1$ day) and with a maximum separation in the sky of 8$^{\prime}$ (Galama
et al. 1998). More recently, prominent spectral features similar to those detected for
SN~1998bw have been detected in the afterglow of GRB030329 (Stanek et al. 2003; Hjorth et al.
2003). To date, no other such compelling case of GRB-SN association has been detected, and
SN~1998bw, whose brightness and kinematic conditions were exceptional, is considered a typical
``hypernova'', the powerful SN explosion speculated to be at the origin of GRBs (Woosley 1993;
Paczy\'nski 1998; Zhang et al. 2003, and references therein).

The field of SN~1998bw and GRB980425 was observed with the BeppoSAX Narrow Field Instruments
one day, one week and 6 months after the event (Pian et al. 2000).  The most sensitive of
these instruments, the BeppoSAX MECS, had detected two X-ray sources within the BeppoSAX Wide
Field Cameras 8$^{\prime}$-radius error box of GRB980425.  The brighter one (hereafter S1) was
variable and positionally consistent with the SN~1998bw and had been identified with X-ray
emission from the SN.  The fainter source (hereafter S2)  is $\sim 4^{\prime}$ away from
SN~1998bw, and therefore inconsistent with it, within the $1^{\prime}.5$ positional
uncertainty of the MECS detectors (see Fig. 1 in Galama et al.  1999). No firm statement could
be made about the variability of S2, given its low flux level, at the limit of the MECS
sensitivity.

Since the X-ray light curve of S2 may have been marginally compatible with an afterglow
behavior, given the big uncertainties (see Pian et al. 2000), some reservations had remained
as to whether GRB980425 and SN~1998bw were physically associated.  However, the above
mentioned case of GRB030329 and SN~2003dh argues strongly in favor of the GRB980425/SN~1998bw
association and considerably weakens the case for association between S2 and GRB980425.

The peculiarity of SN~1998bw made it imperative to further investigate its X-ray emission at
late epochs. Therefore, we have re-observed its field with Chandra and XMM-{\it Newton} in
late 2001 and March 2002, respectively, and report here the preliminary results of the latter
campaign.  A detailed presentation of both campaigns will be given in future papers.

\begin{center}
\begin{figure}
\includegraphics[width=14cm]{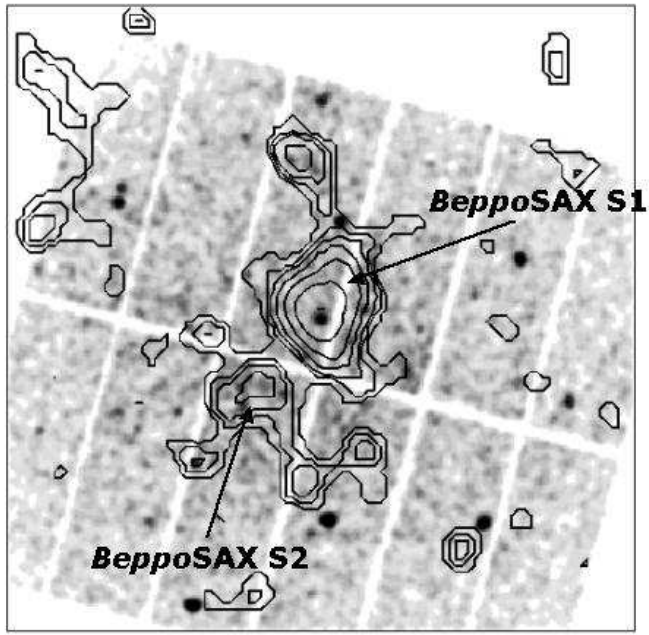}

\caption{XMM-{\it Newton} image of the field of SN~1998bw taken with the EPIC PN chip on 28
March 2002.  The size of the image is $30^{\prime} \times 30^{\prime}$; North is at the top
and East to the left. Overlaid are isophotes of the BeppoSAX MECS image taken in 26-27 April
1998.  The brightest source near the center of the image corresponds to SN~1998bw.  About 4
arcmin SE of it, one can distinguish a number of very faint sources, the sum of which may have
been detected by BeppoSAX as a single source S2 at the limit of MECS detectability.}

\end{figure}
\end{center}

\section*{OBSERVATIONS AND DATA ANALYSIS}

The field of SN~1998bw was observed by XMM-{\it Newton} on 28 March 2002, between 13:53:34 and
20:10:38 UT, with the European Photon Imaging Cameras (EPIC, 0.15-15 keV)  PN (Str\"uder et
al. 2001)  and MOS (Turner et al.  2001), operating in full-frame mode and with the medium
filter applied.  The data have been cleaned and processed using the Science Analysis Software
(SAS 5.3) and analyzed using standard software packages (FTOOLS 5.2). The latest calibration
files released by the EPIC team have been used (update: 29 Jan 2003). Event files produced
from the pipeline have been filtered from high-background time intervals and only events
corresponding to pattern 0-12 for MOS and pattern 0-4 for PN have been used (see the XMM-{\it
Newton} Users' Handbook, Ehle et al. 2001; see also ~~ {\tt
http://xmm.vilspa.esa.es/external}). The net exposure times, after data cleaning, are $\sim
13.0$ ks, $\sim 16.8$ ks, and $\sim 16.7$ ks for PN, MOS1, and MOS2, respectively.  Count
rates (see next Section) have been estimated by integrating the signal within circles of
30$^{\prime\prime}$ radius, which enclose $\sim$80\% of the encircled energy function, and by
subtracting the background signal estimated from blank sky exposures, in circles of equal
area.

\begin{center}
\begin{figure}
\includegraphics[width=12cm]{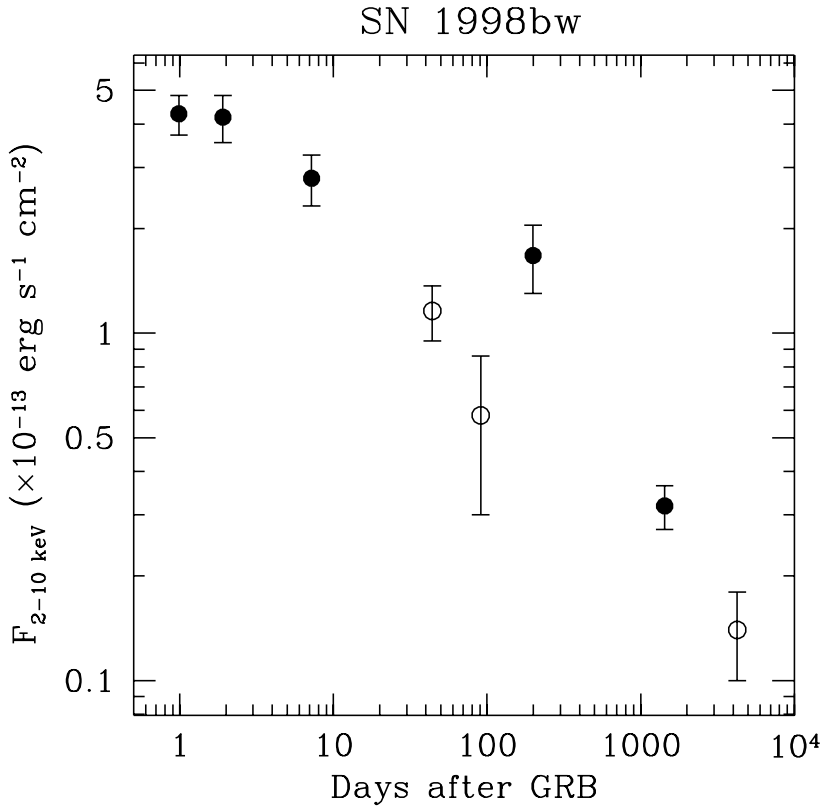} 

\caption{X-ray light curve of SN~1998bw constructed from BeppoSAX MECS (first 4 filled
circles, from Pian et al. 2000) and XMM-{\it Newton} EPIC PN (last filled circle)
observations.  The temporal origin coincides with the trigger time of GRB980425, 1998 April
25.9091 UT. For comparison, the X-ray light curve of SN~1980K is also shown (open circles,
from Canizares et al. 1982; Schlegel 1994. The original data points have been converted to
the 2-10 keV range, using a power-law spectrum $\propto \nu^{-\alpha}$ with index $\alpha \sim
1$).}

\end{figure}
\end{center}

\section*{RESULTS AND DISCUSSION}

In Fig. 1 we show the EPIC PN image.  At a position consistent with that of source S1 in the
BeppoSAX MECS image, we detect in the EPIC PN image a source with a count rate of (9.6 $\pm$
1.4) $\times 10^{-3}$ counts s$^{-1}$ in the 0.5-5 keV range within a circular area of
30$^{\prime\prime}$ radius. Assuming a spectrum similar to that measured by BeppoSAX for S1
(see Pian et al. 2000) and accounting for Galactic absorption by neutral hydrogen ($N_{HI} =
3.95 \times 10^{20}$ cm$^{-2}$, Dickey \& Lockman 1990), we derive a flux of (3.18 $\pm$ 0.46)
$\times 10^{-14}$ erg s$^{-1}$ cm$^{-2}$ between 2 and 10 keV. The signal-to-noise ratio of
our XMM-{\it Newton} observations is not sufficient to perform a detailed spectral analysis.

A number of very faint sources are visible at the location of source S2 detected by the
BeppoSAX MECS in the WFC error box of GRB980425 (Fig. 1). The integral of the EPIC PN signal
within the $1^{\prime}.5$ radius MECS error circle of S2 is consistent with the average flux
measured by the MECS for S2, suggesting that it may be not a single source, but rather the sum
of several faint sources and that its marginal variability in the BeppoSAX observations is
determined by background fluctuations, or possibly by the random variations of those sources.
This definitely rules out the afterglow nature of S2.

From the EPIC imaging we do not detect significant contamination of the X-ray emission of
SN~1998bw by its host galaxy, therefore the SN decay measured by BeppoSAX must be authentic.
That was satisfactorily fitted both by a power-law $t^{-0.2}$ and by an exponential with
$e$-folding time of $\sim$500 days (Pian et al. 2000).  The addition of the EPIC point to the
X-ray light curve of SN~1998bw (Fig. 2) shows that neither a power-law nor an exponential law
are particularly satisfactory, although the exponential may be somewhat better.  For this
reason we favor an interpretation of the X-ray light curve as a result of the superposition of
different radiation components. In fact, while the overall temporal behavior of SN~1998bw is
unlike that of the very few X-ray SNe monitored at both early and late times (e.g.,
SN~1987A,
Park et al. 2002;  SN~1993J, Kohmura et al. 1994; Swartz et al. 2003; Zimmermann \& Aschenbach
2003; SN~1994I, Immler et al. 1998; Immler et al. 2002), at late epochs (i.e., after day
$\sim$100)
it is reminiscent of that of previous X-ray SNe (e.g., SN~1980K, Canizares et al. 1982;
Schlegel 1994; Schlegel 1995; SN~1994I, Immler et al. 2002).  Thus, one may interpret the
early X-rays of SN~1998bw as afterglow radiation, while the late epoch X-ray emission is
dominated by the interaction of the SN shock with the circumstellar material, as proposed for
other X-ray SNe (Kohmura et al. 1994; Schlegel 1995; Suzuki and Nomoto 1995; Fransson et al.
1996; Chevalier and Fransson 2002; Immler and Lewin 2002).  In fact, if the main sequence
progenitor of SN~1998bw was a star of $\sim 40 M_\odot$ as postulated by Iwamoto et al. (1998)
one would expect circumstellar material in the neighbourhood of the SN as a result of an
earlier mass loss wind.

On the other hand, assuming that a single mechanism is responsible for X-ray emission of
SN~1998bw, we may not exclude that the observed X-rays are cooling radiation from the compact
remnant, provided the GRB has swept up all the surrounding material by creating an evacuated
cone. Tavani (1997) has shown, in the context of X-ray afterglows of GRBs, that cooling
neutron stars with ``external'' disturbances (e.g., a fallback) may radiate in X-rays with a
temporal rate faster than a power-law.  A longer diffusion time than considered by Tavani
(1997) should be adopted in our case.  The predicted decay rates for neutron stars with simple
cooling (i.e., no fallback) are much longer than the fading time scale measured by BeppoSAX
and XMM-{\it Newton} for SN~1998bw (Page 1998). However, exotic cooling mechanisms can
significantly increase the cooling rate (Slane et al. 2002). Our observations may thus allow
us to place constraints on non-standard neutron star cooling scenarios and may have important
implications for determining the nature of GRB remnants.

\section*{ACKNOWLEDGEMENTS}

We are grateful to Matteo Guainazzi and Matthias Ehle for their assistance with observations
scheduling, and to Paolo Mazzali for his valuable comments.

\end{document}